# Improving BLE Based Localization Accuracy Using Proximity Sensors

Authors version

Marcin Kolakowski

Institute of Radioelectronics and Multimedia Technology, Warsaw University of Technology, Warsaw, Poland, contact: marcin.kolakowski@pw.edu.pl

# Improving BLE Based Localization Accuracy Using Proximity Sensors

Marcin Kolakowski, *Student Member, IEEE*

*Abstract* — Bluetooth Low Energy systems are one of the most popular solutions used for indoor localization. Unfortunately their accuracy might not be sufficient for some of the applications. One way to reduce localization errors is hybrid positioning, which combines measurement results obtained with different techniques. The paper describes a concept of a hybrid localization system in which Bluetooth Low Energy technology is supported with the use of laser proximity sensors. Results from both system parts are fused using a novel, simple positioning algorithm. The proposed system concept was tested using BLE and proximity sensors evaluation boards.

## I. INTRODUCTION

NARROWBAND localization systems based on technologies such as Bluetooth Low Energy (BLE) or Wi-Fi are among most popular solutions for indoor positioning. Typically, most of them allow for localization with uncertainty of several meters [1] . Such accuracy might not be sufficient in case of some of the possible applications. For example systems used in AAL (Active and Assisted Living) applications for older persons movement trajectory analysis would highly benefit from sub-meter accuracy.

One way to improve localization accuracy are hybrid positioning algorithms employing measurement data obtained with different techniques or independent systems. In the literature various hybrid solutions combining Bluetooth based localization with other techniques can be found.

One of the most popular concepts is fusion of signal power measurement results obtained with BLE modules with data from inertial sensors worn by the user [2]. Solutions employing other narrowband and wideband radio interfaces alongside Bluetooth were also proposed. A description of localization system combining BLE with Wi-Fi is presented in [3]. Paper [4] includes the description of a system utilizing both ultra-wideband (UWB) and BLE. In all of the above works the use of hybrid algorithms allowed for more accurate localization than in case of using only Bluetooth technology.

This work was partly supported by the National Centre for Research and Development, Poland under Grant AAL/Call2016/3/2017 (IONIS project).

Marcin Kolakowski is with the Institute of Radioelectronics and Multimedia Technology, Warsaw University of Technology, Nowowiejska 15/19, 00-665 Warsaw, Poland (phone: +48 22 2347635; e-mail: m.kolakowski@ire.pw.edu.pl).

The following paper describes a concept of a hybrid localization system combining BLE technology with laser proximity sensors. The main application of the system is monitoring everyday life of elderly person living alone. Such systems allows to detect changes in daily routine and emergency situations, which could be helpful in evaluating patient's overall health state. The author has not found any publications regarding the use of proximity sensors in hybrid positioning systems intended for indoor person localization. Typically such sensors are used in robotics for obstacle detection systems [5].

In the proposed concept proximity sensors are an addition to a typical BLE infrastructure. They can be integrated with Bluetooth anchors or can be mounted separately in locations, where propagation conditions may be difficult. The sensors perform continuous ranging at same rate as power measurements in BLE interface. If the localized person finds himself in the sensor vicinity the obtained results are fused with BLE power levels using a novel hybrid positioning algorithm.

The paper describes also an exemplary proximity sensor, that could be used in such system – VL53L1X produced by STMicroelectronics [6]. Sensor's basic features (ranging bias and standard deviation, field of view) were experimentally tested for reflection from a person in various conditions.

The rest of the paper is structured as follows. Section II describes system concept and the algorithm used for localization. Section III includes results of VL53L1X sensor investigation. Results of system concept verification are presented in section IV. Section V concludes the paper.

## II. HYBRID LOCALIZATION SYSTEM CONCEPT

### A. System architecture

The proposed hybrid positioning system architecture is presented in Fig. 1.

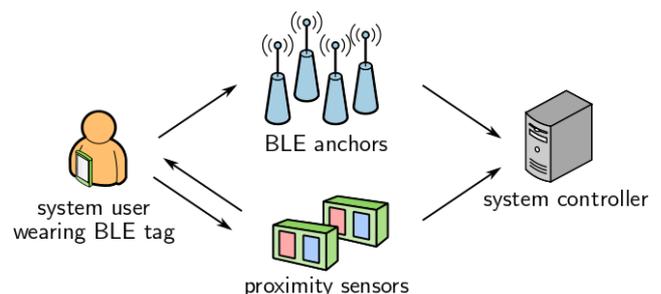

Fig. 1. Proposed system architecture

The system consists of three components: tag, which is worn by the user, system infrastructure comprising several Bluetooth Low Energy anchors and laser proximity sensors and a system controller. The system works the following way. The tag periodically sends BLE packets. BLE anchors measure received signal strength (RSS). At the same time, proximity sensors periodically perform ranging. Measurement results from anchors and lasers are sent to the system controller, where user location is derived using hybrid localization algorithm.

In the proposed functional architecture BLE anchors and proximity sensors are separate devices but they might be integrated in one device. Such approach would simplify communication with the system controller but may make system deployment problematic. Since proximity sensors' ranges are usually limited to a few meters it would be best to orient them so that their beam direction is parallel to the floor and the covered area is as large as possible. In order for their signal to be reflect from the localized person effectively they should be placed at waist to chest level, which from the point of view BLE signal propagation might not be the best possible solution.

### B. Localization algorithm

The results from the anchors and proximity sensors are processed using hybrid localization algorithm, which scheme is presented in Fig. 2

Algorithm work flow consists of two phases. Firstly user localization is calculated based on RSS measurement results obtained with BLE anchors and separately using data from proximity sensors. The obtained results are then fused in the next step.

Received Signal Strength values measured for received Bluetooth signals are processed with Extended Kalman Filter (EKF) based algorithm [4]. In the algorithm the monitored user is described with a state vector containing his coordinates and speed. The algorithm comprises two phases: time update and measurement update phase. In time update phase, current user's location and speed are estimated based on last state vector value and equations of uniform linear motion. In the assumed model user's acceleration is treated as white noise. In measurement update phase, the predicted values are corrected based on RSS measurement results.

Localization using laser proximity sensors is performed only using sensors, which detected user's presence in their vicinity. For each of the sensors, user location is roughly estimated as the point located at the main sensor axis at the measured distance. If locations for two or more proximity sensors are obtained user location is calculated as a weighted average:

$$x = \frac{\sum_i \frac{1}{\sigma_i^2} x_i}{\sum_i \frac{1}{\sigma_i^2}} \quad (1)$$

where $x_i$ is the location derived based for i-th sensor and $\sigma_i^2$ is the variance associated with this result (in this algorithm – variance of distance measurements). Variances of ranging results may vary with distance from the reflecting object and sensor type so their values should be estimated experimentally prior to system deployment.

Assuming independence between all proximity sensors measurements the resulting variance can be estimated as:

$$\sigma^2 = \frac{1}{\sum_i \frac{1}{\sigma_i^2}} \quad (2)$$

The final result of the hybrid algorithm is a weighted average (1) of partial results. In case of BLE based localization the variance is taken from covariance matrix, which is the output of EKF.

### III. VL53L1X SENSOR PROPERTIES

One of the proximity sensors, which could be used in the presented hybrid positioning system concept is VL53L1X produced by STMicroelectronics. The sensor uses Time-of-Flight (ToF) technique for distance measurement and according to the application note [6] allows to perform ranging up to 4 m with millimeter resolution and frequency up to 50 Hz. Sensor's field of view can be programmed in range from 15 to 27 degrees. Another advantage of VL53L1X is that its package is relatively small (4.9x2.5 mm) and therefore it can be easily integrated with localization system circuits.

Before using the sensors to verify system concept, its basic properties such as ranging errors, ranging standard deviation and sensor field of view (FoV) were evaluated. In the experiment P-NUCLEO-53L1A1 [7] modules with SimpleRanging example application were used [8]. The sensor worked in 'long' mode and ranging rate was set to 10 times per second. For the experiment the default FoV of 27 degrees was chosen.

### A. Ranging accuracy and precision

The first part of the experiment consisted in measuring distance from a person dressed in black t-shirt standing directly opposite the sensor. The measurements were performed in an office space on a sunny day with blinds open and then closed to evaluate the impact of ambient light on sensor operation. Additionally to check whether the type of clothing influences ranging performance the experiment was repeated for person clad in a white t-shirt and bare chested. The graphs of ranging error and standard deviation versus distance from the sensor are presented in Fig. 3 and Fig. 4 respectively.

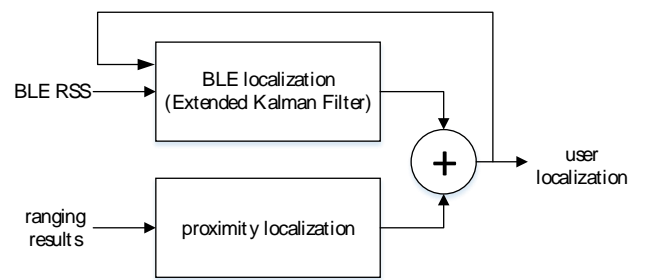

Fig. 2. Algorithm workflow scheme

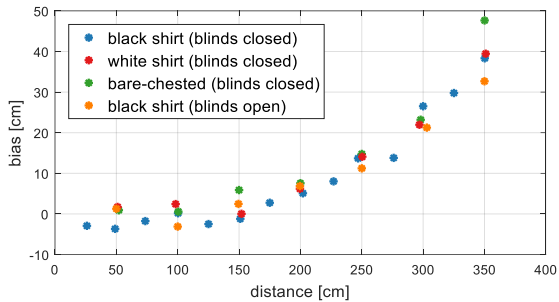

Fig. 3. Ranging bias vs distance

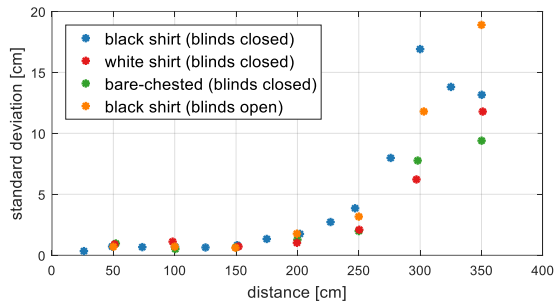

Fig. 4. Ranging standard deviation vs distance

The maximum range for which it was possible to measure distance from the person was about 3.5 m. In case of measurements performed close to the sensor, the ranging bias is small and does not exceed few centimeters. For distances larger than 2 m, the bias rises and for 3.5 m exceeds 30 cm. There is no clear dependence of clothes colors and lightning conditions on ranging bias. It means that in the proposed system it would be possible to use the above results and implement efficient ranging bias mitigation.

Ranging standard deviation does not exceed 5 cm for the distances below 2.5 m. For larger distances it rises and at the sensor maximum range may near 20 cm. There is a standard deviation dependence on the type of reflective surface. For the black t-shirt, standard deviation at distances larger than 2 m was significantly bigger than in case of white one. The obtained results do not depend on lighting conditions.

The obtained results were used to model the relationship between standard deviation and distance from the sensor. It was modeled as a third degree polynomial and used in positioning algorithm described in the previous section.

### B. Sensor FoV evaluation

In the second step of VL53L1X tests, the actual FoV of the sensor was evaluated. The experiment consisted in placing an A3 white sheet attached to a utility cart in front of the sensors and moving it sideways to the point in which the sensor did not measure the distance from the object. Border positions of the cart were noted and based on it sensor's actual FoV was evaluated. The experiment was performed on a sunny day with the blinds opened. The comparison between the evaluated and declared 27° FoV is presented in Fig. 5.

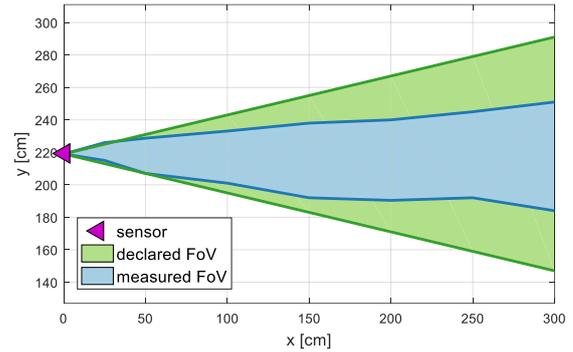

Fig. 5. Measured and declared sensor's field of view

The actual field of view for investigated type of object is smaller than in the application note. From the distance of 50 cm the actual FoV width is significantly smaller and at 3 m the difference is about 80 cm. The observed differences do not disqualify VL53L1X sensors from use in hybrid positioning systems but should be taken into account in localization algorithm and during system deployment.

## IV. EXPERIMENTS

The proposed system concept was experimentally tested in one of Warsaw University of Technology laboratories. The chosen place was demanding one in terms of propagation conditions. The lab was densely cluttered with equipment and in the middle of it stood a table, which was partitioned by a 1.5m high metal board. System infrastructure layout is presented in Fig. 6.

Bluetooth Low Energy subsystem consisted of four Texas Instruments BLE Multi-standard CC2650 LaunchPad evaluation boards [9] acting as anchors and CC2540 USB Evaluation Kit [10] used as a tag. In the experiment direction of BLE transmission was reversed - the anchors transmitted BLE packets 10 times per second.

As the proximity sensors two P-NUCLEO-53L1A1 modules including VL53L1X sensors were used. The modules were placed at the height of 1.4 m, so that the beam was directed at person's corpus.

During the experiment a person was moving along a trajectory consisting of three straight lines. He was localized both with EKF algorithm using BLE results and the algorithm proposed in the paper. In case of the hybrid algorithm, ranging variance at fusion step was calculated based upon the model created in Section III. Localization results are presented in Fig. 6. Positioning accuracy was evaluated based on the distance of localized points from the reference trajectory. The CDFs (cumulative distribution functions) of this feature estimated jointly for both paths are presented in Fig. 7.

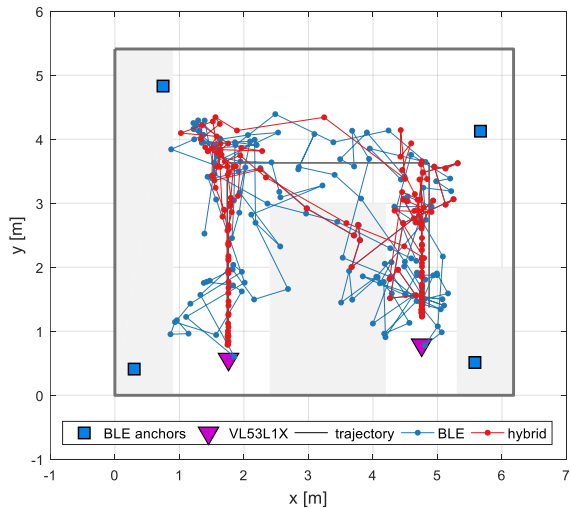

Fig. 6. Localization results. Gray rectangles mark tables located in the lab

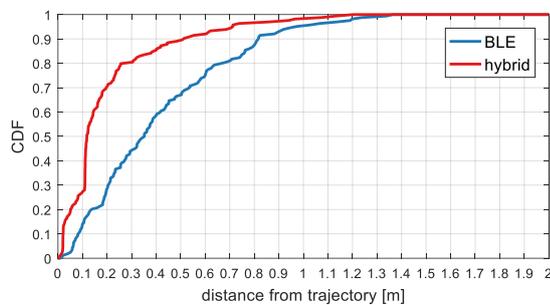

Fig. 7. Estimated CDFs for distance from reference trajectory

Localizations calculated solely upon BLE power measurement results are not accurate enough to properly evaluate user's movement trajectory. Significant fluctuations of measured BLE signal levels result in inaccurate localization, which is most noticeable at the path ends and turns. In those places localization results are mixed up and it is hard to tell, whether the localized person is moving or turning around in place. The additional use of proximity sensors placed in the room allowed to improve localization accuracy. In sensor's vicinity it is possible to precisely determine user's location and detect his movements.

For half of BLE based localization results distance from the reference trajectory exceeds 35 cm. Locations calculated with the proposed hybrid algorithms are closer and median trajectory error is about dozen centimeters.

## V. CONCLUSION

The paper describes a concept of hybrid localization system combining BLE technology with laser proximity sensors. Measurement results from both parts are fused using a novel algorithm, which takes into account ranging standard deviation.

The concept was experimentally tested using BLE radio modules and VL53L1X proximity sensors. Experiments have shown that ranging results returned by the sensors include distance dependent bias. Additionally standard deviation of distance measures rises with the distance and depends on the type of reflective surface. The experiments have also shown that for this type of objects sensor's field of view is smaller than declared in the application note.

The results of localization obtained using hybrid algorithm are more accurate than those obtained only with BLE results.

The presented concept will be furtherly developed by modifying localization algorithm so that it will take sensor's FoV into consideration.